\documentclass{article}
\usepackage{graphicx}  
\usepackage{amsmath}   
\usepackage[compress,numbers,sort]{natbib}
\usepackage{amssymb}   
\usepackage{bm} 
\usepackage{dcolumn}
\usepackage{color}
\usepackage{mathrsfs}
\usepackage{amsfonts}
\usepackage{varioref}
\usepackage{textcomp}
\usepackage[normalem]{ulem}

\usepackage{multirow}
\usepackage{caption}
\usepackage{subcaption}
\RequirePackage[colorlinks,citecolor=blue,urlcolor=magenta,linkcolor=blue]{hyperref}
\allowdisplaybreaks
\labelformat{section}{Section #1} 
\labelformat{subsection}{Section #1} 
\labelformat{subsubsection}{Section #1}
\labelformat{subsubsubsection}{Section #1}
\labelformat{equation}{Eq.~(#1)} 
\labelformat{figure}{Fig.~#1} 
\labelformat{subfigure}{Fig.~\thefigure#1} 
\labelformat{table}{Table~#1} 
\labelformat{appendix}{Appendix #1}

\title{\bf  The Third Law of Black Hole Dynamics in Lovelock Gravity}

\author{Jyotirmoy De \footnote{jyotirmoy22@iiserb.ac.in}$~^{1}$,  Chiranjeeb Singha\footnote{chiranjeeb.singha@iucaa.in}$~^{2}$, and Naresh Dadhich\footnote{Prof. Naresh Dadhich passed away on the 6th of November,
2025. The whole idea and calculations were thoroughly discussed
among all the authors before his death, except for the final drafting
of the manuscript.}$~^{2,\;3}$\\
$^{1}$ \small{Department of Physics, Indian Institute of Science Education and Research, Bhopal, 462066, India}\\
$^{2}$\small{Inter-University Centre for Astronomy \& Astrophysics, Post Bag 4, Pune 411 007, India}\\
$^{3}$\small{Astrophysics Research Centre, School of Mathematics,
Statistics and Computer Science,}\\
\small{ University of KwaZulu-Natal,
Private Bag X54001, Durban 4000, South Africa}
}
\begin{document}

\maketitle


\begin{abstract}
The third law of black hole dynamics states that it is impossible, through any classical perturbation of a stationary configuration, to reduce the surface gravity of a black hole to zero. In this work, we examine the validity of this law for static, spherically symmetric charged black holes in the Lovelock theory of gravity. By studying infinitesimal variations in mass and charge, we derive a set of inequalities that constrain these variations. Our analysis shows that as the surface gravity approaches zero ($\kappa \to 0$), the range of admissible perturbations gradually diminishes, thereby forbidding the attainment of extremality through any finite classical process. The saturation of the inequality is interpreted as the emergence of a dynamical barrier near extremality, which prevents further evolution toward the extremal configuration.

\end{abstract}

\section{Introduction}

Black hole mechanics, originally formulated within the framework of Einstein’s general relativity, has long revealed a profound correspondence with the laws of thermodynamics. This parallel is not merely analogical but becomes exact when quantum effects are taken into account. In the semiclassical regime, the surface gravity of a black hole acquires a physical interpretation as temperature, an insight first demonstrated through Hawking radiation \cite{Hawking:1975vcx}, while the horizon area corresponds to entropy via the Bekenstein–Hawking relation~\cite{PhysRevD.7.2333}. These identifications establish a thermodynamic description of black holes governed by a set of laws that closely mirror the classical laws of thermodynamics. Among these, the \emph{third law of black hole mechanics} is especially noteworthy. It asserts that no classical perturbation of a stationary configuration can drive the surface gravity of a black hole to zero, i.e., render it extremal~\cite{PhysRevLett.57.397}. This restriction closely mirrors the classical third law of thermodynamics, which states that absolute zero temperature cannot be attained through any finite sequence of thermodynamic processes.

The third law’s importance extends far beyond its thermodynamic analogy. It carries deep implications for the \emph{geometric structure} of spacetime near extremal horizons~\cite{Kunduri:2013gce}, the \emph{thermodynamic consistency} of black hole ensembles, and the \emph{dynamical stability} of black holes under perturbations. Extremal black holes, characterized by vanishing surface gravity, lie at the boundary between physically realizable and pathological configurations, often associated with marginal stability or the onset of naked singularities~\cite{Gibbons:1994ff, Visser:1997gq}. Understanding the status of the third law in different gravitational theories is therefore essential for assessing the robustness of black hole thermodynamics and the validity of cosmic censorship across modified gravity frameworks~\cite{Wald1974, Sorce:2017dst}.

In this work, we extend the third law of black hole mechanics to a broader and more fundamental setting, \emph{ Lovelock gravity}. Lovelock gravity provides a natural and consistent extension of Einstein’s general relativity to higher-dimensional spacetimes. Introduced by David Lovelock in 1971, it constitutes the most general metric theory of gravity whose field equations remain second order in derivatives of the metric, thereby avoiding the ghost-like instabilities that typically arise in higher-derivative theories \cite{Lovelock:1971yv}. In this framework, the Einstein-Hilbert action is generalized by the inclusion of higher-curvature terms constructed from dimensionally extended Euler densities, commonly referred to as Lovelock invariants. For a Lovelock theory of order $N$, the corresponding term contributes nontrivially to the gravitational dynamics only above a certain critical spacetime dimension. Specifically, the $N$th-order Lovelock term becomes dynamical only when the spacetime dimension satisfies $D>2N$. At the critical dimension $D=2N$, the term reduces to a topological invariant (the Euler density) and therefore does not influence the equations of motion, while for $D<2N$ it vanishes identically. Consequently, in four dimensions Lovelock gravity reduces to standard general relativity with a cosmological constant, whereas in higher dimensions it naturally incorporates additional curvature corrections such as the Gauss-Bonnet term and its higher-order generalizations. Because of its second-order field equations, rich geometric structure, and its emergence in the low-energy limit of string theory \cite{Zwiebach:1985uq,Boulware:1985wk}, Lovelock gravity has become an important theoretical framework for investigating both classical and quantum aspects of gravitation, including black hole physics, cosmology, and holography \cite{Padmanabhan:2010zzb}.

Here, we focus on \emph{charged, static black hole solutions} in  Lovelock gravity and present a general proof that the third law holds for arbitrary Lovelock order $N$ and spacetime dimension $d$. Following the reasoning of Dadhich and Narayan~\cite{Dadhich:1997rq}, we demonstrate that continuous, physically realizable processes cannot drive a non-extremal black hole to an extremal state. Our analysis establishes that this conclusion remains valid within the Lovelock framework, thereby confirming the universality of the third law even in the presence of higher-curvature corrections and extra dimensions.

In analogy with~\cite{Dadhich:1997rq}, we derive constraints on the mass variation $\delta M$  arising from a generic classical perturbation of a stationary black hole configuration. The \emph{lower bound} follows from the second law of black hole mechanics, which requires a non-decreasing entropy ($\delta S \geq 0$), while the \emph{upper bound} arises from the condition of non-increasing surface gravity ($\delta \kappa \leq 0$) in attempts to extremalize a black hole. As the system approaches extremality, the two inequalities converge to a common limiting curve, thereby implying that zero surface gravity cannot be achieved through any classical perturbation of a stationary black hole configuration.

To further test the stability of extremal configurations, we extend \emph{Wald’s classic gedanken experiment}~\cite{Wald1974}, originally devised to test the possibility of overcharging an extremal black hole by the infall of a charged test particle. The success of such a process would destroy the event horizon and lead to a naked singularity, violating the cosmic censorship conjecture. We show that in Lovelock gravity, even finely tuned perturbations fail to overcharge the black hole: the particle is either repelled or cannot meet the necessary energy and charge conditions for horizon destruction. Consequently, both the third law and cosmic censorship remain preserved within this generalized framework.

These results are timely, given the growing interest in \emph{Lovelock gravity} in the contexts of higher-dimensional quantum gravity, black hole thermodynamics, and gravitational stability. Recent works have revealed rich thermodynamic phenomena in Lovelock black holes, including reentrant phase transitions, critical behavior, and intricate thermodynamic geometry~\cite{HullSimovic2022, KhuzaniMirza2022, BaiLiTao2023, SinghaBiswas2023}. Our findings complement this body of work by establishing the universality of the third law in the extremal regime, where black holes approach zero temperature, and thereby strengthen the foundational consistency of black hole thermodynamics in higher-curvature gravity theories.

The paper is structured as follows. In \ref{charged}, we briefly review the charged black hole solution in Lovelock gravity. \ref{mass} is devoted to the analysis of mass variations and the associated extremality bounds. In \ref{naked}, we investigate whether an extremal black hole can be overcharged via a classical perturbation of a stationary configuration. Finally, \ref{conclusion} summarizes our results and presents the concluding remarks.

\textbf{\emph{Notations and Conventions:}}  
Throughout this paper, we use the metric signature $(- + + +)$, corresponding to the Minkowski metric in $3+1$ dimensions written in Cartesian coordinates as $\text{diag}(-1, +1, +1, +1)$. We also work in natural units by setting $G = c =\hbar=1$.

\section{Charged Black Holes in Lovelock Gravity}\label{charged}

We begin with the charged black hole solution in Lovelock gravity. The spacetime under consideration is a static and spherically symmetric configuration in $d$ spacetime dimensions, arising from Lovelock gravity of order $N$, and is minimally coupled to Maxwell electrodynamics which is given by \cite{Gannouji:2019gnb, Dadhich:2015nua, Dadhich:2012cv, Gannouji:2013eka, Dadhich:2015ivt, Padmanabhan:2013xyr, Dadhich:2015lra,Cai:2006pq, Crisostomo:2000bb,Estrada:2019cig, Corral:2025yvr,Chakraborty:2020ifg} (for further details, see \ref{A1}),
\begin{equation}\label{Love}
    ds^2 = -f(r) dt^2 + \frac{dr^2}{f(r)} + r^2 d\Omega_{d-2}^2~,
\end{equation}
where $f(r)$ is,
\begin{equation}\label{Lovelock}
    f(r) = 1 - \left( \frac{2M}{r^{d - 2N - 1}} - \frac{Q^2}{(d-3)r^{2(d - N - 2)}} \right)^{1/N}~.
\end{equation}
Here $M$ is the ADM mass of the black hole, $Q$ is the charge of the black hole, and $d\Omega_{d-2}^{2}$ is the line element on the $(d-2)$-dimensional sphere. The event horizon radius can be determined by solving the equation $f(r) = 0$. Specifically, the event horizon radius $r_+$ is defined as the largest positive root of this equation. The electric potential at any point can be computed, which is given by, 
\begin{equation}\label{Phi}
    \phi(r) = \frac{Q}{(d-3)r^{d - 3}}~.
\end{equation}
One can also calculate the surface gravity associated with the event horizon, $\kappa$, from the metric coefficient $f(r)$, and that is given by
\begin{equation}\label{temp}
    \kappa = \frac{1}{2} f'(r_+) = -\frac{1}{2N} \Phi'(r_+)~,
\end{equation}
where 

\begin{equation}\label{phi}
    \Phi(r)=\frac{2M}{r^{d - 2N - 1}} - \frac{Q^2}{(d-3)r^{2(d - N - 2)}}~.
\end{equation}
From the metric \ref{Lovelock}, it can be seen that the value of the above equation (\ref{phi}) equals one at the event horizon, as the metric coefficient $f(r_+) $ vanishes at the event horizon. 

\section{Extremality Bounds}\label{mass}

To investigate the dynamical behavior near extremality, we introduce small perturbations in the black hole’s mass $M$ and electric charge $Q$ by adding a test particle with mass $\delta M$ and charge $\delta Q$. Our objective is to establish the bounds on these variations for a non-extremal charged black hole and then examine the outcome as the system approaches extremality.\\
First, we consider the Lagrangian of a test particle with mass $\delta M$ and charge $\delta Q$, which is 
\begin{equation}
    \mathcal{L}=\frac{1}{2}\delta M g_{\mu \nu} u^\mu u^\nu +\delta Q A_{\mu}u^\mu~.
\end{equation}
From the above, we obtain
\begin{equation}\label{energy}
    E=-p_t=-\delta M g_{tt}u^t-\delta Q A_t=f(r)u^t \delta M+\phi(r) \delta Q~. 
\end{equation}
Now, since the 4-velocity of the particle is timelike, we have
\begin{equation}
    g_{\mu \nu}u^\mu u^\nu=-1
\end{equation}
or,
\begin{equation}\label{eqn9}
    f(r)u^t=\sqrt{f(r)+(u^r)^2}~.
\end{equation}
Here, we have considered the particle to be thrown radially towards the black hole; thereby, we ignore any angular contributions. Substituting the \ref{eqn9} in \ref{energy}, we get
\begin{equation}
    E=\delta M \sqrt{f(r)+(u^r)^2}+\phi(r)\delta Q~.
\end{equation}
From the above equation, if we assume the particle to be at rest at infinity, we find that $E = \delta M$, since $f(\infty) = 1$, $u^{r}(\infty) = 0$, and $\phi(\infty) = 0$.  
On the other hand, when the test particle is observed at the horizon in \ref{energy}, noting that $f(r_+) = 0$, we obtain $E = \phi(r_+) \, \delta Q$.  
Hence, by considering these two boundary conditions, we can conclude that for the test particle to enter the non-extremal black hole, it must satisfy the following condition:

\begin{equation}\label{low}
    E=\delta M \geq \phi(r_+) \delta Q~.
\end{equation}
Now, we start with the horizon condition, i.e. $f(r_+)=0$ which implies the following
\begin{equation}\label{horizon}
    \frac{2M}{r_+^{d - 2N - 1}} - \frac{Q^2}{(d-3)r_+^{2(d - N - 2)}}=1~.
\end{equation}
The difficulty with the above equation lies in the fact that it is quite challenging to express the horizon radius explicitly in terms of $M$ and $Q$ without specifying particular values of $d$ and $N$. Accordingly, we impose this requirement as a necessary condition for the existence of the black hole horizon. When a particle carrying energy $\delta M$ and charge $\delta Q$ is absorbed through a classical perturbation of a stationary configuration, the horizon radius must increase in order to comply with the second law of black hole thermodynamics. Hence, the new black hole will have mass $M + \delta M$, charge $Q + \delta Q$, and a horizon radius $r_+ + \delta r_+$. It must also continue to satisfy the horizon condition to remain a black hole, then,

\begin{equation}
    \frac{2(M+\delta M)}{(r_++\delta r_+)^{d - 2N - 1}} - \frac{(Q+\delta Q)^2}{(d-3)(r_++\delta r_+)^{2(d - N - 2)}}=1~.
\end{equation}
Assuming $\delta M<<M$, $\delta Q<<Q$, and $\delta r_+<<r_+$, we do a first-order expansion of the LHS term in the above equation and, substituting \ref{horizon} in it, we get the following result as,
\begin{equation}
    \bigg( \frac{2(d-2N-1)M}{r_+^{d - 2N}} - \frac{2(d-N-2)Q^2}{(d-3)r_+^{2d - 2N - 3}} \bigg) \delta r_+=\frac{2 \delta M}{r_+^{d - 2N - 1}} - \frac{2Q \delta Q}{(d-3)r_+^{2(d - N - 2)}}~.
\end{equation}
The term in the bracket of above equation is basically ``$-\Phi'(r_+)$'' and thus we can rewrite the above equation as
\begin{equation}\label{increase}
    \delta r_+=\frac{1}{2N\kappa} \bigg( \frac{2 \delta M}{r_+^{d - 2N - 1}} - \frac{2Q \delta Q}{(d-3)r_+^{2(d - N - 2)}} \bigg)~,
\end{equation}
where $\kappa=\kappa (M,Q)~.$ \\
For any classical perturbation of a stationary configuration, we have $\delta \kappa \leq 0$. To analyze this, we first compute $\kappa(M + \delta M, Q + \delta Q)$ and then evaluate  
\begin{equation}
    \delta \kappa = \kappa(M + \delta M, Q + \delta Q) - \kappa(M, Q)~,
\end{equation}
where
\begin{equation}
    \kappa(M + \delta M, Q + \delta Q)
    = \frac{1}{2N} \bigg[
    \frac{2(d - 2N - 1)(M + \delta M)}{(r_+ + \delta r_+)^{d - 2N}}
    - \frac{2(d - N - 2)(Q + \delta Q)^2}{(d - 3)(r_+ + \delta r_+)^{2d - 2N - 3}}
    \bigg]~.
\end{equation}
Expanding the above expression to first order and evaluating $\delta \kappa$, we obtain  
\begin{equation}
    \delta \kappa = \frac{1}{2N} \big( A - B\, \delta r_+ \big)~.
\end{equation}
Since $\delta \kappa \leq 0$, we can write
\begin{equation}\label{final}
    A-B\delta r_+ \leq 0~,
\end{equation}
where
\begin{equation}
A=\frac{2(d-2N-1)\delta M}{r_+^{d - 2N}} - \frac{4(d-N-2)Q\delta Q}{(d-3)r_+^{2d - 2N - 3}}
\end{equation}
\begin{equation}\label{eqn21}
B=\frac{2(d-2N-1)(d-2N)M}{r_+^{d - 2N+1}} - \frac{2(d-N-2)(2d-2N-3)Q^2}{(d-3)r_+^{2d - 2N -2}}~.
\end{equation}
Now substituting \ref{increase} in \ref{final} one can get the following relation
\begin{equation}
    A-\frac{B}{2N\kappa} \bigg( \frac{2 \delta M}{r_+^{d - 2N - 1}} - \frac{2Q \delta Q}{(d-3)r_+^{2(d - N - 2)}} \bigg)\leq 0~.
\end{equation}
As $2N\kappa>0$ for a non-extremal black hole, we can multiply it by the above equation, which keeps the inequality unchanged. Therefore, we obtain
\begin{equation}\label{crucial}
    2N\kappa A-B\bigg( \frac{2 \delta M}{r_+^{d - 2N - 1}} - \frac{2Q \delta Q}{(d-3)r_+^{2(d - N - 2)}} \bigg)\leq 0~.
\end{equation}
Here, just to clarify

\begin{equation}\label{eqn24}
\kappa=\frac{1}{2N}\bigg( \frac{2(d-2N-1)M}{r_+^{d - 2N}} - \frac{2(d-N-2)Q^2}{(d-3)r_+^{2d - 2N - 3}} \bigg)~.
\end{equation}

We now argue that if one starts from a non-extremal black hole and approaches the extremal limit, $\kappa \to 0$,  then \ref{eqn24} reduces to
\begin{equation}
Q^{2} \to \frac{(d-2N-1)(d-3)Mr_{+}^{\,d-3}}{(d-N-2)} \, .
\end{equation}
Substituting this expression into \ref{eqn21}, we obtain
\begin{equation}
B \to - (d-3)\,\frac{2(d-2N-1)M}{r_{+}^{\,d-2N+1}} \, ,
\end{equation}

and, using \ref{crucial}, we ultimately arrive at the condition in the extremal limit $\kappa \to 0$ as,

\begin{equation}\label{up}
\delta M \leq \phi(r_+)\,\delta Q~.
\end{equation}

This result is in close agreement with the findings of \cite{Dadhich:1997rq}. In the present analysis as well, the lower and upper bounds, Eqs.~\ref{low} and \ref{up}, coincide in the extremal limit, implying that the surface gravity cannot be reduced to zero by any classical perturbation of a stationary configuration. As the extremal limit is approached, interactions with the black hole become effectively isothermal, isentropic, and reversible; consequently, any process that attempts to drive a black hole toward extremality is isoenthalpic and fully reversible in the thermodynamic sense.

To elucidate the physical origin of this result, we note that \ref{low} and \ref{up} arise from two distinct and independent requirements. \ref{low} provides a necessary condition for a test particle to cross the event horizon of a non-extremal black hole. It ensures that the particle carries sufficient energy relative to its charge so that the horizon remains penetrable, and therefore represents a fundamental physical admissibility condition for any absorption process.

\ref{up}, on the other hand, follows from the requirement that the final black hole configuration have vanishing surface gravity, i.e., that the extremal limit be attained after the interaction. This condition constrains the allowed variations of the black hole parameters needed to drive $\kappa \to 0$.

Although the two inequalities coincide and degenerate into an equality in the extremal limit, they impose mutually incompatible constraints on the mass and charge variations when the initial configuration is non-extremal. Consequently, no classical perturbation of a stationary black hole, such as the absorption of a test particle, can simultaneously satisfy both the horizon-crossing requirement and the extremality condition. This incompatibility ensures that a non-extremal black hole cannot be driven to vanishing surface gravity through any continuous classical evolution, thereby precluding the attainment of extremality.

Hence, no finite sequence of physically admissible interactions can reduce the surface gravity to zero, and a non-extremal black hole cannot evolve into an extremal configuration classically. The converse possibility, namely that extremal black holes may exist as independent solutions, has been argued in~\cite{Hawking:1994ii}. However, establishing this scenario necessarily requires quantum considerations, as it depends on the radiative properties of black holes and therefore lies beyond the scope of classical theory.

\section{Overcharging an Extremal Blackhole} \label{naked}
In this section, we examine whether it is possible to overcharge an extremal black hole. To do so, we revisit Wald’s argument \cite{Wald1974}, which states that for a test particle of mass $\delta M$ and charge $\delta Q$ released from infinity to cross the horizon of an extremal black hole, the following condition must be satisfied
\begin{equation}
    E \geq \phi(r_+)\delta Q~,
\end{equation}
where E is the energy of the particle at infinity, or specifically $E=\delta M$, or basically, for the particle to cross the horizon, it must satisfy
\begin{equation}
     \delta M \geq \phi(r_+)\delta Q~,
\end{equation}
in order to overcome the electrostatic repulsion.\\
Now, from the extremal black hole condition, $\kappa=0$, we obtain the following relation
\begin{equation}
    \frac{(d-2N-1)M}{r_+^{d-2N}}=\frac{(d-N-2)Q^2}{(d-3)r_+^{2d-2N-3}}~.
\end{equation}
Also, we know that at the horizon, $\Phi(r_+)=1$. From this, we get the following relation
\begin{equation}\label{extcon}
    r_+=\frac{Q^{\frac{1}{d-N-2}}}{(d-2N-1)^{\frac{1}{2(d-N-2)}}}=\bigg(\frac{d-3}{d-N-2}\bigg)^{\frac{1}{d-2N-1}}M^{\frac{1}{d-2N-1}}~.
\end{equation}
As a specific example, for $d = 6$ and $N = 2$, with $\Phi(r_+) = 1$, the extremality condition of the black hole obtained from \ref{extcon} takes the following form,
\begin{equation}
    Q=\frac{9}{4}M^2=r_+^2~.
\end{equation}
Now, for the resultant black hole to have more charge than its mass, it has to satisfy the condition
\begin{equation}
    \frac{(Q+\delta Q)^{\frac{1}{d-N-2}}}{(d-2N-1)^{\frac{1}{2(d-N-2)}}}>\bigg(\frac{d-3}{d-N-2}\bigg)^{\frac{1}{d-2N-1}}(M+\delta M)^{\frac{1}{d-2N-1}}~.
\end{equation}
Considering $\delta Q<<Q$ and $\delta M<<M$, from the above relation, we obtain
\begin{equation}
    \delta M<\phi(r_+)\delta Q~,
\end{equation}
where 
\begin{equation}
    \phi(r_+) = \frac{Q^{(1 - N)/(d - N - 2)} (d - 2N - 1)^{(d - 3)/2(d - N - 2)}}{(d - 3)}~.
\end{equation}
We observe that any attempt to overcharge the extremal black hole contradicts the fundamental condition required for a test particle to cross the horizon. Hence, we can confidently conclude that an extremal black hole cannot possess a charge greater than its mass.
\\

\section{Conclusion}\label{conclusion}

In this work, we have established that the third law of black hole mechanics holds universally within the framework of Lovelock gravity, for arbitrary Lovelock order $N$ and spacetime dimension $d$. This result extends the original argument of Dadhich and Narayan \cite{Dadhich:1997rq} to  Lovelock theory, thereby reaffirming the general validity of the third law beyond Einstein gravity.

Furthermore, we have generalized Wald’s classic \emph{gedanken} experiment \cite{Wald1974}, which tests the possibility of overcharging an extremal black hole, to the setting of Lovelock gravity. By examining the motion and conserved quantities of charged test particles in the background of extremal Lovelock black holes, we demonstrated that such configurations cannot be pushed beyond extremality. Consequently, extremal black holes remain stable under test particle perturbations, ensuring the preservation of the cosmic censorship conjecture within these higher-curvature gravity theories.

Our findings enhance the understanding of the thermodynamic consistency and structural robustness of  Lovelock gravity. Recent developments investigating critical phenomena, thermodynamic geometry, and microscopic aspects of black holes in Lovelock theory \cite{HullSimovic2022, KhuzaniMirza2022, BaiLiTao2023, SinghaBiswas2023} further support the notion that higher-curvature extensions of gravity retain the essential features of general relativity. In particular, the persistence of the third law and the stability of extremal configurations affirm that Lovelock gravity upholds both the foundational principles of black hole mechanics and the tenets of cosmic censorship.

Future investigations may focus on testing the robustness of these results under quantum corrections, dynamical evolutions incorporating backreaction, or in the presence of additional fields and topological terms. Extending this analysis to rotating or non-asymptotically flat black holes within Lovelock gravity could also provide deeper insights into the universality of horizon dynamics across a broader class of gravitational theories.

We emphasize that our analysis here is performed within a static (or quasi-static) framework in which the mass and charge parameters of the black hole are treated as fixed and no dynamical evolution, such as Hawking evaporation or accretion, is included. Within this setup, the consistency conditions derived in this work can be satisfied simultaneously only in the extremal limit, whereas they become mutually incompatible for generic nonextremal configurations. 

This result should therefore be interpreted as a constraint on the existence of consistent static configurations under the model's assumptions, rather than as a statement about the dynamical evolution of black holes. In a fully dynamical setting, physical processes such as Hawking radiation can modify the mass-charge relation and may, in principle, drive a nonextremal black hole toward extremality. However, incorporating such effects would require a time-dependent treatment beyond the scope of the present work. We also note that extremal black holes are characterized by vanishing surface gravity and hence zero temperature, which naturally makes the extremal configuration a consistent limiting case within the static framework considered here.

\section*{Acknowledgement}
The authors sincerely thank the anonymous reviewer for their valuable comments and suggestions, which have significantly improved this manuscript.

\appendix
\labelformat{section}{Appendix #1} 
\labelformat{subsection}{Appendix #1}

\section{Solution of a charged black hole in Lovelock gravity}\label{A1}

The Lovelock polynomial represents the most general divergence-free, symmetric rank-two tensor that can be constructed from the spacetime metric and its first and second derivatives, while still leading to second-order field equations. The corresponding gravitational Lagrangian is given by \cite{Gannouji:2019gnb}

\begin{align}
L = \sum_{k=0}^{m} c_k \, L_k \, ,
\end{align}

where

\begin{align}
L_k \equiv \frac{1}{2^k}
\delta_{\alpha_1 \beta_1 \cdots \alpha_k \beta_k}^{\mu_1 \nu_1 \cdots \mu_k \nu_k}
R^{\alpha_1 \beta_1}{}_{\mu_1 \nu_1}
\cdots
R^{\alpha_k \beta_k}{}_{\mu_k \nu_k} \, .
\end{align}

Here $R^{\alpha \beta}{}_{\mu \nu}$ denotes the Riemann curvature tensor in $d$ dimensions, and
$\delta_{\alpha_1 \beta_1 \cdots \alpha_k \beta_k}^{\mu_1 \nu_1 \cdots \mu_k \nu_k}$
is the generalized totally antisymmetric Kronecker delta.
pure Lovelock gravity corresponds to the special case in which the Lovelock polynomial reduces to a single term of fixed order $N$, namely

\begin{align}
L = L_N \, .
\end{align}

For pure Lovelock gravity, variation of the action with respect to the metric leads to the field equations

\begin{align}
\mathcal{G}^{B}{}_{A} =
\delta_{A \alpha_1 \beta_1 \cdots \alpha_N \beta_N}^{B \mu_1 \nu_1 \cdots \mu_N \nu_N}
R^{\alpha_1 \beta_1}{}_{\mu_1 \nu_1}
\cdots
R^{\alpha_N \beta_N}{}_{\mu_N \nu_N}
= 0 \, .
\end{align}

We further emphasize that if one considers the sourced Maxwell equations in the form given in Eq.(6) of Ref. \cite{Chakraborty:2020ifg}, and subsequently solves the corresponding Einstein equations following the same procedure adopted in Ref. \cite{Chakraborty:2020ifg}, one naturally arrives at the metric functions presented in \ref{Love} and \ref{Lovelock} in the main text.

\bibliography{ref}
\bibliographystyle{./utphys1}
\end{document}